\documentclass[10pt,conference]{IEEEtran}
\ifCLASSINFOpdf
   \usepackage[pdftex]{graphicx}
   \DeclareGraphicsExtensions{.pdf,.jpeg,.png}
\else
   \usepackage[dvips]{graphicx}
   \DeclareGraphicsExtensions{.eps}
\fi
\usepackage{amsmath}
\usepackage{amssymb}
\usepackage{amsthm}
\usepackage{cancel}
\usepackage{breqn}
\usepackage{multirow}
\usepackage{rotating}
\usepackage{cite}
\usepackage{verbatim}

\newcommand{\norm}[1]{\left\lVert#1\right\rVert}
\begin{document}
%
\title{Dual Polarized Modulation and Receivers for Mobile Communications in Urban Areas}
%
%
%
\author{\IEEEauthorblockN{Pol Henarejos\IEEEauthorrefmark{1}}
\IEEEauthorblockA{\IEEEauthorrefmark{1}Centre Tecnol\`{o}gic de Telecomunicacions de Catalunya\\Castelldefels, Barcelona\\Email: pol.henarejos@cttc.es}
\and
\IEEEauthorblockN{Ana Isabel P\'{e}rez-Neira\IEEEauthorrefmark{1}\IEEEauthorrefmark{2}}
\IEEEauthorblockA{\IEEEauthorrefmark{2}Universitat Polit\`{e}cnica de Catalunya\\Barcelona\\
Email: ana.isabel.perez@upc.edu}
}

\maketitle

\begin{abstract}
Achieving an increase in the spectral efficiency (SE) has always been a major driver in the design of communication systems. The use of MIMO techniques in mobile communications has achieved significant benefits in improving the system throughput. The basic underlying concept of MIMO is to exploit the signal and channel characteristics to eliminate interference between multiple transmissions. Departing from the work carried out under the industrial projects \cite{NGW,Henarejos}, we extend the results provided in their respective reports. The goal is to increase the SE without an increment of radiated energy without any Channel State Information at Transmitter (CSIT) and feedback at the transmitter and maintaining a very low computational complexity at the receiver. Although a priori additional power is required to increase the SE, we demonstrate that the proposed Polarized Modulation (PMod) scheme exploits the polarization diversity reducing the required EbN0 and adding an extra bit. We also introduce two receivers to demodulate this scheme of different computational complexities. We describe a Near Optimal Detector (NOD) which achieves almost the same performance as the optimal detector based on the Maximum Likelihood Detection (MLD), but with lower computational complexity. Finally, the results demonstrate that the PMod requires less EbN0 compared with the single polarization case, guarantees the robustness in the presence of the cross-polarization and validates that PMod can multiplex two streams of different Quality of Service (QoS).
\end{abstract}


%
\IEEEpeerreviewmaketitle

\section{Introduction}
%
%
%
%
\IEEEPARstart{T}{he} increase of spectral efficiency (SE) has been a major line of research in the present and past centuries. Multiple-input multiple-output (MIMO) schemes appeared to boost the SE by the addition of antennas at the transmitter and receiver \cite{SatelliteCommun.2008,Schwarz2008,Electr.amp2011} without increasing the radiated energy. Even though the SE can be enhanced using Channel State Information at the Transmitter (CSIT), it requires an additional feedback channel, which is not always possible. 

Vertical Bell Laboratories Layered Space-Time (VBLAST) scheme and successive improvements are schemes that do not require CSIT and increase the SE with a relative increase of the processing complexity \cite{Wolniansky1998,Golden1999,Shen2003,Xue2003}. However, VBLAST introduces interference between the streams since all the signals are transmitted through all antennas without any interference pre-cancellation. Based on VBLAST, recent research projects such as \cite{NGW,NICOLE,&Technol.Centre2010}, reported that the polarization of the antennas can increase the SE. The major drawback is that the complexity of the terminals increases due to the interference across the polarizations. 

In contrast to the previous works, where the polarization multiplexing is used mainly for the increase of the SE, the concept of space-time codes is also applied to polarization-time domains (PTC). In \cite{Frigyes2005}, a basic concept of PTC is applied and extended in \cite{Sellathurai2006} for different scenarios. A measurement campaign and a validation of the model is provided in \cite{Horvath2007}.

In conjunction to VBLAST and PTC, we introduce the Polarized Modulation (PMod), which exploits the polarization diversity and increases the SE, and propose two receivers. In the PMod, a single stream is placed in one polarization depending on an extra bit. Thus, the election of the polarization is in fact another dimension to carry information. PMod complements the polarization shift keying (PolSK) firstly appeared in the optical communications. In fibre cables, although the polarization is not preserved, the offsets and drifts in the polarization planes are perfectly controlled. This fact motivated the appearance of the PolSK \cite{Benedetto1992}. The concept is translated in the free-space in some works studying the applicability \cite{Tang2010,Wang2014} and some works doing measurements \cite{Soorat2012}. Additionally, other works propose new polarization shapes, where the information is modulated in these shapes \cite{Leong2005}, and an additional polarized modulation employing the phase is presented in \cite{Wei2013}.

However, the main difference between PolSK and PMod is the fact that in PMod a symbol is also transmitted, whereas in PolSK the information is placed exclusively in the polarization state. Hence, in terms of SE, PMod offers more degrees of freedom since it combines the polarization dimension and the amplitude of in-phase and quadrature of the transmitted symbol. Although the literature in this field is scarce or even non-existent, our research concluded that using PMod consumes less power compared with the single polarization case to achieve the same Bit Error Rate (BER) carrying extra information.

Finally, works such as \cite{Asplund2007,Coldrey2008} suggest that the vertical and horizontal polarizations (V/H), in average terms, are preserved in an urban scenario. Although it is well known that the propagation of V/H varies from the shapes of the environment, in an urban scenario walls, floor and roofs predominate in the same manner. Thus, on average, the V/H are preserved and therefore the use of polarization as transmission scheme is encouraged. As we demonstrate in the next sections, PMod becomes the favourite trade-off between the robustness of the PTC schemes and the higher SE schemes such as VBLAST.

\section{System Model}
Consider a point-to-point MIMO system with dual polarized antennas at transmission and reception, able to transmit and receive in two orthogonal polarizations at the same time. The channel is modelled as a typical Rayleigh one for urban areas. Using a single channel access, one symbol $s$ is transmitted, where $b$ bits are modulated using the constellation $\mathcal{S}$.
We also aim to transmit the symbol $s$ using one of the polarizations depending on an extra bit $c$. If the two polarizations are numbered as $0$ and $1$, then the symbol $s$ is conveyed through polarization $0$ if $c=0$ or through polarization $1$ if $c=1$. Hence, assuming that the channel remains invariant during the channel use and is flat over the system bandwidth, we formulate the system model as
\begin{equation}
\mathbf{y}=\mathbf{Hx}+{\boldsymbol\omega}
\end{equation}
where $\mathbf{y}\in\mathbb{C}^2$ is the received signal, $\mathbf{H}\in\mathbb{C}^{2\times2}$ is the channel matrix, $\mathbf{x}\in\mathbb{C}^2$ is the conveyed signal and ${\boldsymbol\omega}\in\mathbb{C}^2$ is the noise contribution which is a bivariate normal distribution $\mathcal{N}\left(\mathbf{0},\boldsymbol\Sigma\right)$. Additionally, $\mathbf{x}$ is coded as $\mathbf{x}=\left(\begin{smallmatrix}1-c\\c\end{smallmatrix}\right)s$. It is important to remark that although the polarizations may be orthogonal (they do not interfere with each other), the characteristics of the channel may break this assumption and correlate the signals. Hence, even though the symbol $s$ is transmitted in one polarization, the symbol is received in both polarizations at the receiver side. As we explained in the introduction, this feature may be exploited since it provides more diversity at reception.

With this approach, the symbol $s$ is transmitted through two different channels corresponding to the two different polarizations. If the channel matrix is decomposed as $\mathbf{H}=(\mathbf{h}_0\, \mathbf{h}_1)$, the system model can be simplified as in the Single-Input Multiple-Output (SIMO) case:
\begin{equation}
\mathbf{y}=\mathbf{h}_cs+\boldsymbol\omega.
\end{equation}

\subsection{PMod as Hierarchical Modulation}
The total SE of the PMod is given by the SE of the transmission of symbol $s$ plus the extra bit conveyed through the switching polarization step. Hence, we define the SE Gain in terms of the relative gain between PMod and SISO schemes as
\begin{equation}
G=\frac{SE+1}{SE}=1+\frac{1}{SE}.
\end{equation}

It is shown that the gain is higher when the SE is low. This motivates the use of PMod in low order modulations, especially used in low Signal-to-Noise Ratio (SNR) regimes.

In fact, PMod is the superposition of a binary phase shift keying (BPSK) plus the modulation used by symbol $s$. Thus, this scheme can be envisaged as a hierarchical modulation (HM). HM is interesting for its flexibility and CSIT is not necessary. The receiver is able to decode the lowest modulation order (bit $c$) and, if the SNR is higher than a threshold, it can also recover the information of higher modulation order ($b$ bits). 

Hence, in the PMod scheme, the receiver can obtain the extra bit $c$ and recover the bits $b$ if the SNR is high enough. This concept can be used for queue prioritization. Whilst the high priority queues (HPQ) are often composed of lower modulations in order to be decoded in poor SNR margins, the low priority queues (LPQ) offer higher throughputs even though they require higher SNR values.

\begin{figure}[!ht]
	\centering
    \includegraphics[width=0.5\linewidth]{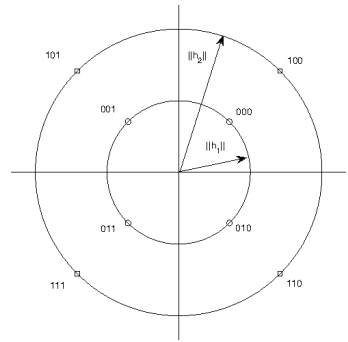}
    \caption{Hierarchical Constellation transformed by PMod}
    \label{fig:const}
\end{figure}

As the transmitted signal is conveyed with PMod, the receiver sees the constellation $\mathcal{S}$ in two copies with radius $\norm{\mathbf{h}_0}$ and $\norm{\mathbf{h}1}$ respectively. For example, if $\mathcal{S}$ is a QPSK constellation, PMod transforms $\mathcal{S}$ as fig.  \ref{fig:const} depicts.

Following this example, deciding between the inner or the outer circles determines the Most Significant Bit (MSB). Then, the rest of Less Significant Bits (LSB) can be decoded. It is clear that HPQ only decides between inner and outer circles, whereas LPQ decides among the quarters.

It is important to recall that PMod uses less power to achieve the same BER compared with the single polarization case and also achieves higher SE, as it is showed in the results section. This is motivated by the fact that using two polarizations increases the diversity and reduces the BER; and an extra bit is appended to the transmission due to the polarization switching.

\section{Detection Schemes}

In this section we describe two approaches to demodulate PMod.

\subsection{Optimal Detector}

The optimal detector is described by the Maximum Likelihood Detector (MLD). The rule of detection is based on the following
\begin{equation}
\hat{\mathbf{x}}=\arg\min_{\mathbf{x}}\norm{\mathbf{y}-\mathbf{Hx}}.
\label{eq:mld}
\end{equation}

Particularly, in the PMod context, the transmit vector $\mathbf{x}$ is restricted to the subset $\mathbf{x}\in\left\{\left(\begin{smallmatrix}s\\0\end{smallmatrix}\right),\left(\begin{smallmatrix}0\\s\end{smallmatrix}\right)\,\left|\,s\in\mathcal{S}\right.\right\}$. Therefore, the receiver performs an exhaustive search within this set and obtains the candidate that minimizes (\ref{eq:mld}). 

After that, $s$ and $c$ can be detected straightforwardly. Hence, if $\mathbf{x}=\left(\begin{smallmatrix}s\\0\end{smallmatrix}\right)$, then $c=0$; if $\mathbf{x}=\left(\begin{smallmatrix}0\\s\end{smallmatrix}\right)$, then $c=1$. In any case, $s$ can be detected using a common SISO receiver.

However, this scheme presents a notable increase of computational complexity. Traditionally, the MLD approach requires an exhaustive search, whose computational complexity may be prohibitive. On the contrary, due to the restriction of the set aforementioned, the cardinality of the search space is reduced from the initial space $ \mathcal{O}\left(2^{b^2}\right)$ to $\mathcal{O}\left(2^{b+1}\right)$, which is equivalent to the complexity of the SISO case where $b+1$ bits are conveyed. 

From the previous paragraph, we are able to conclude that the PMod does not increase the complexity of the existing schemes. More important, the scheme can be integrated into existing SISO receivers as it does not require to process the transmitted signal as a vector but as a scalar.

\subsection{Near Optimal Detector}
In order to decrease the complexity of the receiver to avoid the need to perform an exhaustive search, we address the MLD receiver through the Likelihood Ratio. This scheme is suboptimal but as the results unveil in the next section, it is very close to the MLD curve and therefore it is a Near Optimal Detector (NOD). Without loss of generality, we assume $\boldsymbol\Sigma=\mathbf{I}_{2\times2}$. Thus, decomposing the soft bit decoding, the bit $c$ is obtained from the following rule:
\begin{equation}
\Lambda\left(\mathbf{y}\right)=\frac{P_1^c}{P_0^c}=\frac{P\left(c=1|\mathbf{y}\right)}{P\left(c=0|\mathbf{y}\right)}=\frac{\sum\limits_{\tilde{s}\in\mathcal{S}}\exp\left(-\|\mathbf{y}-\mathbf{h}_1\tilde{s}\|^2\right)}{\sum\limits_{\tilde{s}\in\mathcal{S}}\exp\left(-\|\mathbf{y}-\mathbf{h}_0\tilde{s}\|^2\right)}.
\label{eq:lr_def}
\end{equation}

\begin{figure*}[!t]
\normalsize
\begin{equation}
r=\frac{1}{P_0^c+P_1^c}\left(P_0^cy_0+P_1^cy_1\right)=\frac{y_0\sum\limits_{\tilde{s}\in\mathcal{S}}\exp\left(-\|\mathbf{y}-\mathbf{h}_0\tilde{s}\|^2\right)+y_1\sum\limits_{\tilde{s}\in\mathcal{S}}\exp\left(-\|\mathbf{y}-\mathbf{h}_1\tilde{s}\|^2\right)}{\sum\limits_{\tilde{s}\in\mathcal{S}}\exp\left(-\|\mathbf{y}-\mathbf{h}_0\tilde{s}\|^2\right)+\sum\limits_{\tilde{s}\in\mathcal{S}}\exp\left(-\|\mathbf{y}-\mathbf{h}_1\tilde{s}\|^2\right)}
\label{eq:sd}
\end{equation}
\hrulefill
\vspace*{4pt}
\end{figure*}

This rule applies for all PMod schemes, independently of the chosen constellation $\mathcal{S}$. This presents an advantage since it can be decoupled from the current implementations. Since bit $c$ is independent of the others, it may be decoded separately from the pre-existing decoding chains.

After the detection of bit $c$, the symbol $s$ can be recovered from the signal received at polarization $c$. In other words, it is possible to decode the symbol $s$ using the signal $y_c$. Though, on the one hand, as we mentioned in the introduction, the channel breaks the orthogonality of the polarizations; on the other hand, this fact provides higher diversity. In other words, the symbol $s$ is received in both polarizations and it can be exploited to increase the overall performance.

The intuitive approach to exploit the diversity is to take the arithmetic mean at each polarization. Nevertheless, soft information contained in (\ref{eq:lr_def}) can be used to weigh the signals of each polarization. Therefore we propose to use the signal $r$ in (\ref{eq:sd}) to detect the symbol $s$.

\section{Results}
In this section we analyse and compare the performance of the MLD and the NOD implementations. For the purpose of this section, a QPSK is used. Channel realizations follow  Rayleigh statistic and include fast and slow fading. We consider the system as narrowband and therefore the channel's magnitude remains almost flat in the frequency domain. To emulate the mobility we introduced a Doppler factor of $4$ Hz. The polarization isolation is considered $26.215$dB and it is a common value from electronic components. For simplicity, we assume perfect channel estimation and synchronization.

In all results the following labels are used:
\begin{itemize}
\item\emph{PMod MLD}: PMod with the Maximum Likelihood Detector.
\item\emph{PMod NOD}: PMod with Near Optimal Detector.
\item\emph{Single}: scenario where a single and static polarization is used (basic case).
\item\emph{OSTBC}: corresponds to the Orthogonal Space Time Block Codes applied to polarization-time diversity rather than space-time \cite{Perez-Neira2008}.
\item\emph{VBLAST}: polarization multiplexing with VBLAST scheme. The detector is based on the Minimum Mean Square Error (MMSE).
\end{itemize}

\begin{figure}[!ht]
  \centering
    \includegraphics[width=1\linewidth]{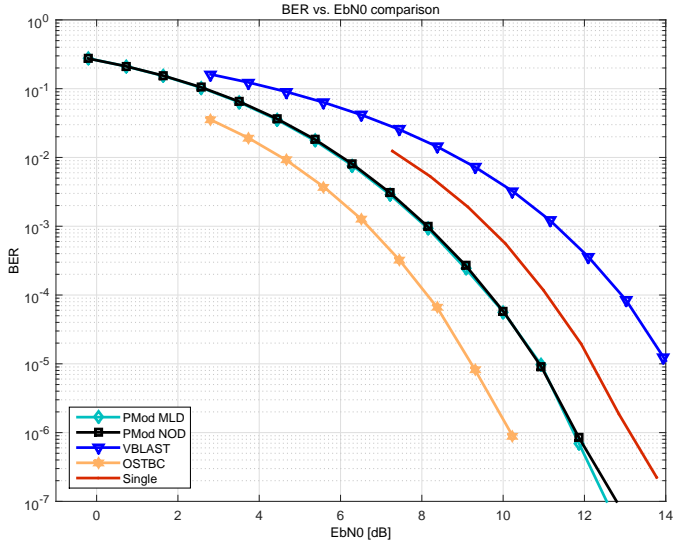}
    \caption{BER comparison of the existing schemes with the MLD and NOD detectors for PMod}
    \label{fig:BER_VTC}
\end{figure}

Fig. \ref{fig:BER_VTC} describes the BER for different values of EbN0 and for different schemes. As we introduced in the previous section, the MLD outperforms the PMod NOD detection. However, the difference between MLD and NOD is less than $0.3$dB and the complexity of NOD is drastically reduced. The negligible difference in BER justifies the name NOD for this detector.

When compared with the single polarization scenario, it is particularly striking that although the single polarization scenario has less SE, it achieves higher BER. On the contrary, PMod carries more information and achieves low BER values. Hence, we are able to conclude that the PMod obtains better results: lower BER and higher SE.

OSTBC obtains the best results as expected. An Alamouti precoder in the polarization-time domain is applied in transmission and a MLD in reception. In contrast to PMod, OSTBC exploits the polarization and temporal diversities. Although it outperforms PMod by less than $2$dB, the SE is not increased. Thus, with an increment of less than $2$dB it is possible to obtain the same error levels but with an increment of $50\%$ of SE using the PMod scheme.

Finally, VBLAST obtains the worst error rates. With this scheme each polarization interferes to each other. Although the receiver applies an MMSE filter, it underperforms PMod.

\begin{figure}[!ht]
  \centering
    \includegraphics[width=1\linewidth]{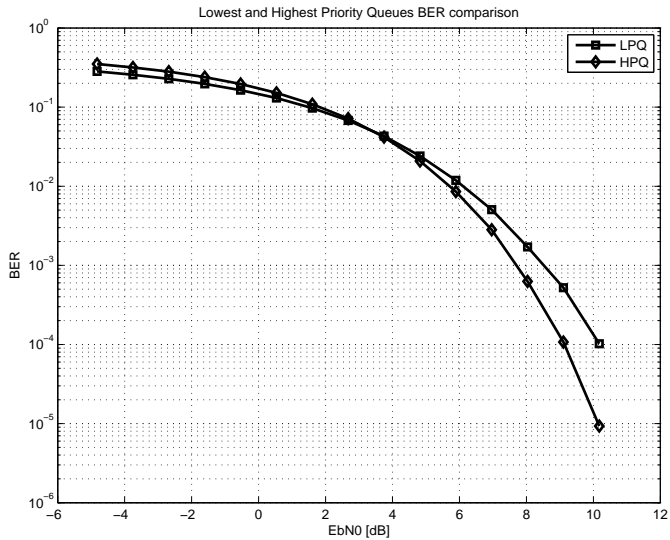}
    \caption{BER comparison between LPQ and HPQ}
    \label{fig:Stream_BER_VTC}
\end{figure}

Fig. \ref{fig:Stream_BER_VTC} illustrates the BER for each PMod queue (LPQ and HPQ). The HPQ is conveyed using the polarization bit $c$ whereas the LPQ is transmitted through the constellation $\mathcal{S}$ carried by the symbol $s$. In this case, HPQ requires less EbN0 to achieve the same level of BER compared with LPQ. Hence, the PMod scheme can be envisaged as a hierarchical modulation as it provides different QoS.

Since PMod offers different grades of BER, it is possible to multiplex different communication services depending on their requirements in terms of QoS. As mentioned in the previous section, we multiplex the LPQ in the curve that obtains higher BER since it carries more bits and has lower QoS requirements. On the contrary, HPQ is devoted on the most restrictive curve. Hence, although it conveys less bits than LPQ, the requirements are more exigent and satisfies the highest priority.

To analyse the impact of the cross-polarization discrimination (XPD), we study the Block Success Rate (BLSR), defined as $BLSR=1-BLER$, where $BLER$ is the Block Error Rate (BLER). With this metric, we are able to measure the effectiveness of the scheme in the presence of XPD. 

Prior to this, the XPD is defined as
\begin{equation}
XPD=20\log\left(\frac{\left|y_c\right|}{\left|y_{1-c}\right|}\right),
\end{equation}
where $y_c$ is the amplitude of the signal received at the polarization where the symbol is transmitted and $y_{1-c}$ is the other one.

\begin{figure}[!ht]
  \centering
    \includegraphics[width=1\linewidth]{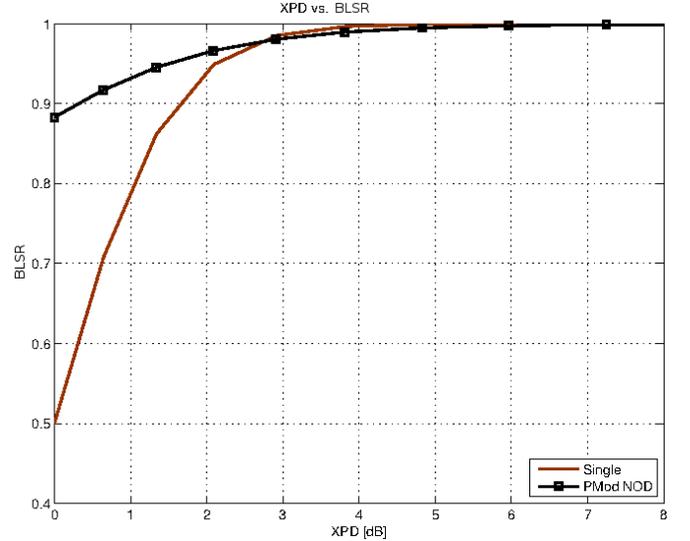}
    \caption{Gain Degradation in the presence of XPD}
    \label{fig:XPD_VTC}
\end{figure}

Examining fig. \ref{fig:XPD_VTC} we can observe the gain degradation (adimensional) for different values of XPD. 

Fig. \ref{fig:XPD_VTC} reveals that the gain is not degraded in the PMod scheme. However, in the scenario with a single polarization, the gain is reduced to its half. This means that the PMod scheme is more robust in the presence of the cross polarization with respect to the single polarization case. The reason is twofold:
\begin{itemize}
\item For low XPD values, both polarizations carry the same symbol. In this case, the probability of error of detecting the bit $c$ increases as the XPD decreases. Note that the probability of error of detecting the symbol $s$ remains the same. This is due to the fact that, even in the case where the $c$ bit is erroneous and the detected polarization is the wrong one, it also contains the symbol $s$ and is able to detect it.
\item For high XPD values, only one polarization carries the data symbol whilst the other only carries noise. In this case, the system will detect the symbol $s$ and the switching bit $c$ correctly.
\end{itemize}
In any case, the system is able to detect the symbol $s$ since XPD increases the polarization diversity at the receiver.

\section{Conclusions}
This paper introduces a novel application to mobile communications of the entitled Polarized Modulation. The work describes that the spectral efficiency increases by a factor of $1+b^{-1}$ in the absence of CSIT. The results demonstrated that the PMod consumes less energy to increase the SE and improves the robustness in the presence of cross-polarization when compared with single polarization scenarios. Two receivers are proposed depending on the computational complexity. Although the optimal is based on the Maximum Likelihood Detection, a Near Optimal Detector is also described. The second receiver (NOD) achieves almost the same performance with MLD but with much lower computational complexity. The results reveal that PMod is a trade-off between OSTBC and VBLAST solutions. Finally, PMod scheme demonstrates the enhancement of spectral efficiency and the robustness in the presence of XPD.

\section*{Acknowledgement}
This work has received funding from the European Comission under the project Newcom$\#$ (FP7-ICT-318306), from the Spanish Ministry of Economy and Competitiveness (Ministerio de Economia y Competitividad) under project GRE3N-LINK-MAC TEC2008-06327-C03-01 / TEC2008-06327-C03-02 and from the Catalan Government (2014-SGR-1567).


%

\ifCLASSOPTIONcaptionsoff
  \newpage
\fi



\bibliographystyle{IEEEtran}
%
\bibliography{biblio}

%




\end{document}